\documentclass[twocolumn,showpacs,amsmath,amssymb, nobibnotes, aps, prl,showkeys]{revtex4}

\usepackage{graphicx}
\usepackage{dcolumn}
\usepackage{bm}
\usepackage{docs}

\expandafter\ifx\csname package@font\endcsname\relax\else
 \expandafter\expandafter
 \expandafter\usepackage
 \expandafter\expandafter
 \expandafter{\csname package@font\endcsname}%
\fi

\begin{document}

\title{Implications of supernova remnant origin model of galactic cosmic rays on Gamma rays from young supernova remnants}

\author{Prabir Banik\thanks{Email address: pbanik74@yahoo.com} and Arunava Bhadra\thanks{Email address: aru\_bhadra@yahoo.com}}

\affiliation{ High Energy $\&$ Cosmic Ray Research Centre, University of North Bengal, Siliguri, West Bengal, India 734013\\}
\begin{abstract}

It is widely believe that galactic cosmic rays are originated in supernova remnants (SNRs) where they are accelerated by diffusive shock acceleration process at supernova blast waves driven by expanding SNRs. In recent theoretical developments of the diffusive shock acceleration theory in SNRs, protons are expected to accelerate in SNRs at least up to the knee energy. If SNRs are true generator of cosmic rays, they should accelerate not only protons but also heavier nuclei with right proportion and the maximum energy of heavier nuclei should be atomic mass (Z) times that of protons. In this work we investigate the implications of acceleration of heavier nuclei in SNRs on energetic gamma rays those are produced in hadronic interaction of cosmic rays with ambient matter. Our findings suggest that the energy conversion efficiency has to be nearly double for the mixed cosmic ray composition instead of pure protons to explain the observation and secondly the gamma ray flux above few tens of TeV would be significantly higher if cosmic rays particles can attain energies Z times of the knee energy in lieu of 200 TeV, as suggested earlier for non-amplified magnetic fields. The two stated maximum energy paradigm will be discriminated in future by the upcoming gamma ray experiments like Cherenkov Telescope array (CTA). 

\end{abstract}

\maketitle

\section{Introduction}

Even after more than a hundred years of their discovery the origin of cosmic rays is not convincingly known. Among the observed features of cosmic rays, the energy spectrum provides significant clues about their origin. The observed energy spectrum of cosmic rays extends from MeV to about 300 EeV energy and is well described by a universal (falling) power law. However, the slope of the energy spectrum changes at least at two points, one around 3 PeV energy where the spectral index steepens from $-2.7$ to $-3.1$ (the so called knee of the spectrum) and another around 3 EeV where the spectrum again flattens to the pre-knee slope (the so called the ankle of the spectrum) [1]. Recent observations also claim for a second knee around 80 PeV [2]. Any viable model of origin of cosmic rays has to explain all these spectral features of energy spectrum.

It is widely believed that bulk of the cosmic rays observed at the Earth, particularly those with energies below the ankle (or below the second knee) are of galactic origin [3]. Among the galactic sources Supernova remnants (SNRs) are considered as the most viable sources of galactic cosmic rays [3,4]. Such a proposition has two strong bases: the energy released in supernova explosions satisfies the energy requirement to maintain cosmic ray energy density considering an overall efficiency of conversion of explosion energy into Cosmic ray particles (hereafter will be termed as \lq conversion efficiency \rq throughout the article) is of the order of $10\%$ [3] and secondly the diffusive shock acceleration operating in SNR can provide the necessary power law spectral shape of accelerated particles with spectral index -2.0 (or slightly less than that) [5] that subsequently steepen to -2.7, as observed, due to energy dependent diffusive propagation effect [3]. 

Some experimental evidences, though circumstantial, have been reported in recent years in favor of SNR origin of Cosmic rays which mainly arise from astronomical studies of SNRs in the gamma ray regime. If SNRs are true generator of cosmic rays, TeV gamma rays being expected to arise from cosmic ray interactions with the ambient matter and the radiation field in the SNRs [6]. Over the last fifteen years or so GeV to TeV gamma rays from a few SNRs have been detected by the modern gamma ray observatories with fluxes consistent with the standard scenario of supernova origin of cosmic rays [7]. Here note that high energy gamma ray fluxes from supernovae also can be explained by the so-called leptonic scenario in which TeV gamma rays produced by inverse Compton scattering of accelerated electrons with diffuse radiation fields. A clear signature of gamma rays originated from pion decay ($\pi^{o} \rightarrow 2\gamma$) should be peaking of the gamma ray emission spectrum at $67.5$ MeV  and evidence for a cutoff below several hundreds of MeV from some SNRs has already been found by AGILE and Fermi [8]. The observation of TeV neutrinos from SNRs will be another clean signature for the hadronic acceleration in supernovae which are not detected yet.   

A few issues of the SNR origin model are, however, are not yet established which include efficiency of conversion of supernova explosion energy to cosmic rays and the maximum energy that can be attained by a cosmic ray particle in SNR. Though there is no firm upper limit of conversion efficiency, a high conversion efficiency is difficult to achieve.     
The other key unsettled issue is the maximum attainable energy. The maximum energy that can be attained by a cosmic ray particle in an ordinary SNR when the remnant is passing through a medium of density $N_{H} \; cm^{-3}$ is [9]
 
\begin{eqnarray}
E_{max}\simeq 4 \times 10^{5} Z \left( \frac{E_{SN}}{ 10^{51} \; erg} \right)^{1/2} \left( \frac{M_{ej}}{ 10 M_{\odot}} \right)^{-1/6} \\ \nonumber
\left( \frac{N_{H}}{ 3 \times 10^{-3} \; cm^{-3}} \right)^{-1/3}\left( \frac{B_{o}}{ 3 \mu G} \right) \; GeV 
\end{eqnarray}
 
which for proton primary is falling short even of the knee of the cosmic ray energy spectrum by about one order of magnitude. The problem is, however, somewhat alleviated by the fact that the effective magnetic field strength at the shock can be amplified due to growth of magnetic waves induced by accelerated cosmic rays.  With the amplified field, the maximum energy achieved in SNR possibly can reach the knee for protons while for Fe nuclei it can reach the second knee.    

While estimating gamma ray contribution from SNRs, usually protons are considered as accelerated particles in SNRs. However, if SNRs are the true sites of cosmic rays, they should also emit other heavier nuclei. Under the context the purpose of the present work is twofold – first we would like to examine the spectral behavior of produced gamma rays and the conversion efficiency in a few SNRs considering that SNRs accelerate cosmic rays with right composition [10]. Secondly we would consider the maximum attainable energy in SNR will be  $Z \times 3 \times 10^{15}$ eV, where $Z$ is the atomic number, as may be achievable under amplified magnetic field scenarios and shall explore the consequences in the secondary gamma ray spectrum. 
 
The organization of the paper is as the following: in the next section we shall describe the methodology for evaluating the TeV gamma rays fluxes generated in interaction of cosmic rays with the ambient matter in SNRs. In section III we shall estimate the hadronically produced GeV-TeV gamma ray fluxes from four young SNRs and one middle aged SNR, those have been detected in GeV to TeV energy ranges, and will compare with the observed spectra. We shall discuss our results in section IV and finally conclude in the same section.

\section{Methodology}

The cosmic rays production spectrum at the shock front of SNR shall follow a power law [5]

\begin{equation}
 \frac{dN}{dE} =K E^{-\alpha} .
\end{equation}
where $K$ denotes the proportionality constant and $\alpha$ is the spectral index. Here $\xi$ is the fraction of the total energy of the supernova explosion E$_{SN}$ transferred to the cosmic ray particles. The observed gamma ray spectra from different SNRs are not always possible to describe in terms of interaction of hadronic cosmic rays with ambient matter if cosmic ray energy spectrum is taken a single power law [7]; instead in some cases a broken power law of SNR accelerated cosmic ray energy spectrum has to consider. For single power law (SPL), the proportionality constant $K$ can be written as [11]

\begin{eqnarray}
K= \frac{(\alpha-2)\xi E_{SN}}{E_{min}^{2-\alpha}-E_{max}^{2-\alpha}} \hspace{1cm} \mbox{if}\hspace{0.5cm} \alpha > 2 \\ \nonumber
 = \frac{\xi E_{SN} }{\ln(E_{max}/m_p c^2)} \hspace{0.6cm} \mbox{if}\hspace{0.5cm} \alpha = 2\;
\end{eqnarray}
where $E_{min}$ is the minimum energy and $E_{max}$ is the maximum energy attainable by a Cosmic ray particle in the SNR.

On the other hand, for broken power law (BPL), the proportionality constant $K$ takes the form
\begin{eqnarray}
\scalebox{0.9}{$
   \begin{aligned}
K= \xi E_{SN} \left[\frac{(E_{b}^{2-\alpha_1} - E_{min}^{2-\alpha_1})}{(2-\alpha_1)}+ \frac{E_{b}^{\alpha_2}}{E_{b}^{\alpha_1}}\frac{(E_{max}^{2-\alpha_2} - E_{b}^{2-\alpha_2})}{(2-\alpha_2)}\right]^{-1}\\ \nonumber 
  \hspace{1cm} \mbox{if}\hspace{0.5cm} E_{CR} \le E_{b} \\ \nonumber 
  = \xi E_{SN} \left[\frac{(E_{b}^{2-\alpha_1} - E_{min}^{2-\alpha_1})}{(2-\alpha_1)}\frac{E_{b}^{\alpha_1}}{E_{b}^{\alpha_2}}+ \frac{(E_{max}^{2-\alpha_2} - E_{b}^{2-\alpha_2})}{(2-\alpha_2)}\right]^{-1}\\ \nonumber 
  \hspace{1cm} \mbox{if}\hspace{0.5cm} E_{CR} > E_{b} \\ \nonumber
   \end{aligned}$} \\
\end{eqnarray}
where $\alpha_1$ and $\alpha_2$ are the spectral indices below and above the break energy $E_b$ of the primary cosmic ray spectrum in SNR respectively and $E_{CR}$ is the energy of an accelerated cosmic ray nuclei.

The shock accelerated cosmic rays interact with the ambient matter (protons) of density $n_{H}$ and produce neutral pions($\pi^{0}$) along with the other particles. The emissivity of so produced $\pi^{0}$ mesons is given by [12,13]

\begin{eqnarray}
Q_{\pi}^{Ap}(E_{\pi}) = c n_{H}\int_{E_{N}^{th}(E_{\pi})}^{E_{N}^{max}}\frac{dn_{A}}{dE_{N}}\frac{d\sigma_{A}}{dE_{\pi}}(E_{\pi},E_{N})dE_{N}
\end{eqnarray}
where $E_{N}^{th}(E_{\pi})$ is the threshold energy per nucleon, determined through kinematic considerations required to produce a pion with energy $E_{\pi}$. Here $d\sigma_{A}/dE_{\pi}$ is the differential inclusive cross section for the production of a pion with energy $E_{\pi}$ in the lab frame by the stated process. We have used the following model with parametrization of the differential cross section for the inclusive cross section as given by [12,14]

\begin{equation}
\frac{d\sigma_{A}}{dE_{\pi}}(E_{\pi},E_{N}) \simeq \frac{\sigma_{0}^{A}}{E_{N}}F_{\pi}(x,E_{N})
\end{equation}
where $x = E_{\pi}/E_{N}$. The inelastic part of the total cross section of p-p interactions ($\sigma_{0}$) is given by [15] 
\begin{eqnarray}
\sigma_{0}(E_{N}) = 34.3+1.88L+0.25L^2 \, mb
\end{eqnarray}
where $L = \ln(E_{N}/TeV)$. We consider two different kind of \lq A \rq dependence, $A^{3/4}$ [16,12] and $A$ [17], in nuclear enhancement factor;  the first one ($A^{3/4}$) approximately takes into account the A-dependence of the inelastic cross section [16] and the later one also considers the fact that only a fraction of projectile nucleons take part in the interaction, not all which leads to overall $A$ enhancement.

We used the empirical function that well describes the results obtained with the SIBYLL code by numerical simulations for the energy distribution of secondary pions as given below [15]  
 
\begin{eqnarray}
F_{\pi}(x,E_{N}) = 4\beta B_{\pi}x^{\beta-1} \left(\frac{1-x^{\beta}}{1+rx^{\beta}(1-x^{\beta})} \right)^{4} \\ \nonumber
\times \left(\frac{1}{1-x^{\beta}}+\frac{r(1-2x^{\beta})}{1+rx^{\beta}(1-x^{\beta})}  \right) \left(1-  \frac{m_{\pi}}{xE_{N}}\right)^{1/2}\;
\end{eqnarray}
where $B_{\pi} = a + 0.25 $, $\beta = 0.98/\sqrt a $, $a = 3.67+0.83L+0.075L^2$, $r=\frac{2.6}{\sqrt{a}}$ and $L = \ln(E_{N}/TeV)$.

The resulting gamma ray emissivity due to decay of $\pi^0$ mesons is given by 

\begin{eqnarray}
Q_{\gamma}^{Ap}(E_{\gamma}) = 2\int_{E_{\pi}^{min}(E_{\gamma})}^{E_{\pi}^{max}}\frac{Q_{\pi^{0}}^{Ap}(E_{\pi})}{(E_{\pi}^2-m_{\pi}^2)^{1/2}}dE_{\pi}
\end{eqnarray}
where the minimum energy of a pion is $E_{\pi}^{min}(E_{\gamma}) = E_{\gamma} + m_{\pi}^2/(4E_{\gamma})$, required to produce a gamma ray photon of energy $E_{\gamma}$ .

The differential flux of gamma rays reaching at the earth, therefore, can be written as   

\begin{eqnarray}
\frac{d\Phi_{\gamma}}{dE_{\gamma}}(E_{\gamma}) = \frac{1}{4\pi D^{2}}Q_{\gamma}^{Ap}(E_{\gamma})
\end{eqnarray}
where $D$ is the distance between the SNR and the Earth. Therefore, if the explosion energy, ambient matter density and distance of a SNR are known, the differential flux from that SNR can be evaluate from the above equation. In the next section we would estimate fluxes of a few SNRs using the above expressions. 

\section{TeV gamma ray fluxes from a few SNRs}

In order to compare the theoretical expectation of high energy gamma rays with the observations at least a few individual SNRs with known values of relevant physical parameters are required which are available at present. We considered a general theoretical framework based on DSA taking the simple one zone model (i.e. the GeV and TeV gamma ray production regions fully overlapped). For the individual SNRs considered here, the parameters like explosion energy, ambient matter density and the distance of the source are known from other considerations. We only choose the spectral index of the SNR accelerated cosmic rays in each individual SNRs so that the derived gamma ray spectrum for the object reasonably matches with the observed spectrum. 

The TeV gamma rays have so far been detected from more than ten shell type SNRs by the Cerenkov telescopes [18]. Nearly half of them are detected by Fermi as well in the GeV energy range. Here we would consider four young shell type SNRs and one middle age SNR which are emitter of gamma rays both in the GeV and TeV energy range. 

Note that the cosmic ray composition at source should differ from the observed abundances [10] due to propagation effect [19]. In fact due to nuclear fragmentation, composition at source is expected to be slightly heavier. A s a first approximation we shall consider the cosmic ray composition in SNR to be the same to the observed cosmic ray composition. We shall further take the same power law index for each of the nuclear species. 

\subsection{Cassiopeia A}

Cassiopeia A (Cas A) is the youngest known supernova remnant of age about $350$ yrs [20] and located at a distance of $~ 3.4$ kpc from the Earth [21]. It is a type IIb supernova from a star of large initial mass, estimated to be between $15$ and $25$ M$_{\odot}$ by the observations of the scattered light echo from the supernova explosion [22]. Cas A is observed in almost all the wavebands, e.g. radio, optical, X-rays and gamma rays (see references in [23]). In GeV gamma rays the source has been observed by  FERMI-LAT [23], whereas the HEGRA [24], MAGIC [25], and VERITAS [26] telescopes detected the source at TeV energies. It is a unique galactic astrophysical source for studying the origin of galactic cosmic rays as well as high-energy phenomena in extreme conditions due to its brightness in different wavelengths. A recent model [27] reproduced the observations of the angle averaged radii and velocities of the forward and reverse shocks and characterized by a total ejected energy $E_{SN} = 2.3 \times 10^{51}$ erg with an envelope mass $M_{env} = 4M_{\odot}$ [27]. X-ray observations predicts that the remnant is currently still interacting with the wind with a post-shock density ranging between $3$ and $5$ cm$^{−3}$ at the current outer radius of the remnant, r$_{SN}$ $\sim2.5$ pc.

The observed GeV–TeV gamma-ray spectrum from CAS A can be explained by hadronic interactions of cosmic rays with the ambient (proton) matter, when a power-law spectrum of protons with a power law index $2.3$ is considered and the maximum energy of cosmic ray protons is taken as $100$ TeV [23]. A harder spectrum with power law index $2.1$ also describes the observed spectrum when an exponential cutoff at 10 TeV is adopted [23].      

We have estimated gamma ray flux produced in the hadronic interaction of cosmic rays with the ambient protons taking cosmic rays with observed mixed composition accelerated in the SNR and considering the maximum attainable energies is up to $Z \times 3 \times 10^{15}$ eV. Our results are shown in figure 1 along with the observed spectrum. It is found that a BPL energy spectrum of accelerated cosmic rays with spectral index $\alpha = -1.7$ below 50 GeV and $\alpha = -2.45$ above 50 GeV reproduces the observed GeV-TeV gamma ray data well by interacting with the ambient matter. The efficiencies of conversion of the supernova explosion energy require $10\%$ for proton and $16.5\%$ and $14\%$ respectively for mixed composition with nuclear enhancement proportional to $A^{3/4}$ and $A$. It is noticed that around and above $100$ TeV the flux is significantly higher when the maximum energy of cosmic ray is taken up to $Z \times 3$ PeV than that due to maximum energy of $200$ TeV case. Future large area telescope like CTA, therefore, should able to probe the maximum attainable energy up to which cosmic ray can accelerate in supernova remnant like CAS A.

\begin{figure}[h]
  \begin{center}
\includegraphics[scale=0.44]{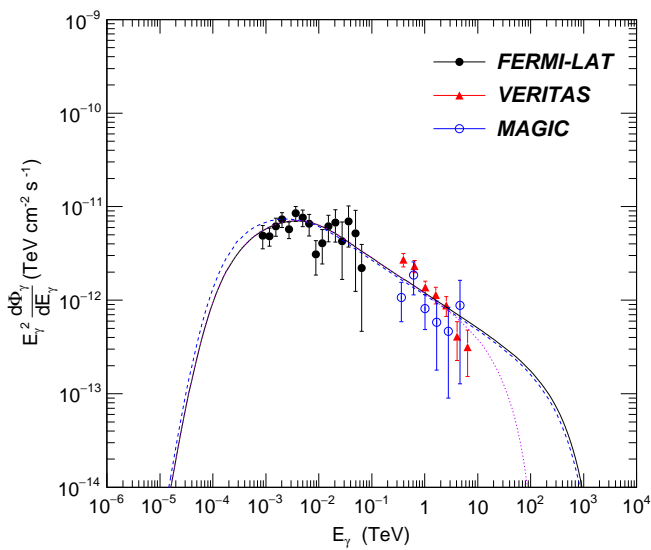}
\end{center}
\caption{Estimated differential energy spectrum of gamma rays reaching at the Earth from Cas A SNR. The black continuous line and blue dashed line denote gamma ray flux for SNR accelerated pure protons and mixed primaries respectively. The pink dash-dotted line denotes the gamma ray flux considering maximum attainable energy of cosmic ray protons as 200 {\bf{TeV}}  }
\label{Fig:1}
\end{figure}

\subsection{Tycho Supernova Remnant}

Tycho's SNR, one of the youngest remnant in the Galaxy, is originated from a Type Ia in $1572$ due to a thermonuclear explosion of a binary system. Fermi has observed Tycho in the GeV energies [28] whereas the VERITAS collaboration observed the source in the $1-10$ TeV range. The observed overall gamma ray spectrum of Tycho is found consistent with the early theoretical predictions [6]. A single power-law with a photon index $2.1$-$2.2$ can describe the GeV-TeV energy spectrum well [29,30] 

We have made the same kind of analysis as we did for Cas A. The distance of the source from the Earth is not very conclusively determined; we have taken the distance as $2.8$ kpc [31]. The density of the ambient matter is $0.9$ cm$^{−3}$ and the explosion energy is $1.2 \times 10^{51}$ ergs [32,33]. We find that a single power law of accelerated cosmic ray energy spectrum with the spectral index $\alpha = -2.3$ describes the GeV-TeV observed gamma rays data well as shown in Fig.2. For proton, $12\%$ efficiency of conversion of supernova explosion energy to cosmic ray energy can explain the experimental results whereas $19.8\%$ and $16.8\%$ conversion efficiency has to be taken for nuclear enhancement by $A^{3/4}$ and $A$ respectively to explain the observations with the mixed primaries.

\begin{figure}[h]
  \begin{center}
\includegraphics[scale=0.44]{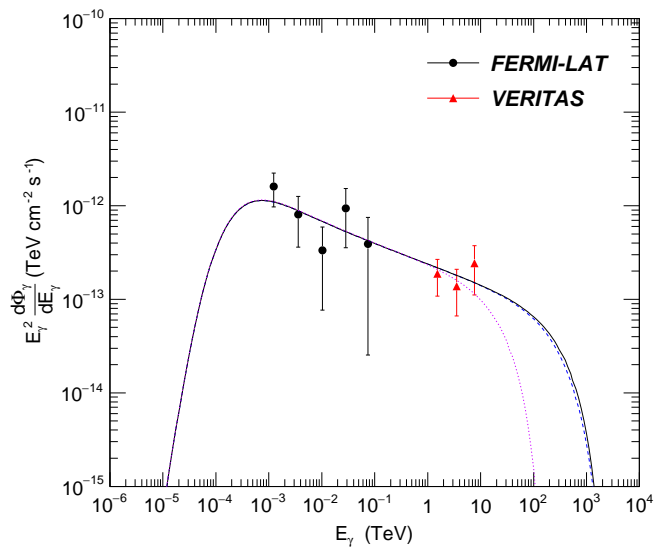}
\end{center}
\caption{Same as Fig.1 but for SNR Tycho}
\label{Fig:2}
\end{figure}

\subsection{SN1006}
The source SN 1006 remnant appeared in the southern sky on 1006 May 1 and was recorded by Chinese and Arab astronomers. In recent years the source has been detected in the GeV gamma ray energy by Fermi [334 and in TeV energies by HESS telescope [35]. The gamma ray flux from SN1006 is, however, quite low, just about {\bf{a$\%$}} of Crab flux. The gamma ray flux is found mainly concentrated in two extended regions, one in North-east and another one in South-west. The observed overall gamma ray spectrum can be interpreted as a consequence of interaction of supernova shock accelerated cosmic rays (protons) with the ambient matter. In such a scenario either the power law spectral index may be taken as 2.3 above 1 TeV with essentially no upper cut-off and a harder spectral index ($<2.0$) below 1 TeV or alternatively a single flat power law with index $\sim 2.0$ with exponential cut-off at 80 TeV. [35].     

A distance of 2.2 kpc of SN1006 was reported by Winkler et al. [36] by comparing the optical proper motion with an estimate of the shock velocity derived from optical thermal line broadening assuming a high Mach number single-fluid shock [35]. We consider the explosion energy  of supernova is $E_{SN} = 2.4 \times 10^{51}$ ergs [37]. The SN 1006 is about 500 pc above the Galactic plane where the external gas density is rather low $n_H = 0.08$ cm$^{-3}$ [35,37]. The estimated differential gamma ray flux reaching at the earth from this SNR is shown in Fig.3 along with the observations by FERMI-LAT and HESS telescopes. The observed data can be explained well considering the spectral index of accelerated cosmic ray spectra in the SNR $\alpha = -2.05$. A significant flux difference has been noticed above 100 TeV between the scenarios with the maximum energy of cosmic rays $3$ PeV and $200$ TeV. When we consider proton as a primary cosmic ray spectrum up to $200$ TeV, the conversion efficiency of 10\% is needed to fit the FERMI-LAT and HESS observational data. Instead if the energy of primary protons is extended  up to 3 PeV, a 11.5\% efficiency of energy conversion is required whereas for the mixed primaries the efficiency has to be taken 16\% and 12.5\% considering nuclear mass enhancement factor $A^{3/4}$ and $A$ respectively. 

\begin{figure}[h]
  \begin{center}
\includegraphics[scale=0.44]{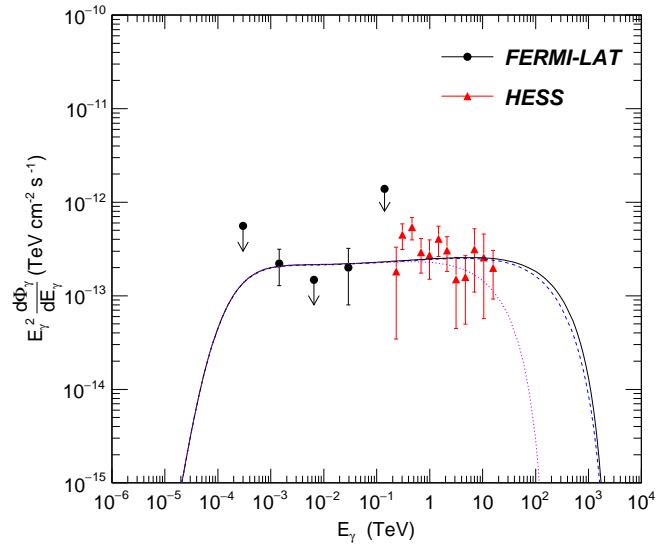}
\end{center}
\caption{Same as Fig.1 but for SNR SN1006. }
\label{Fig:3}
\end{figure}

\subsection{RX J17137-3946}

RX J1713.7–3946 is a young shell-type SNR located in the Galactic plane within the tail of the constellation Scorpius and the age of the object is 1600 years [38]. It is one of the best-studied SNRs from which both non-thermal X-rays and TeV gamma rays are detected. The CANGAROO collaboration in 1998 [38,39] reported first the detection of TeV gamma-ray emission from the SNR and it is confirmed by the subsequent observations with CANGAROO-II in 2000 and 2001 [40]. Later, a resolved image of the source in TeV gamma rays [41] is obtained by the H.E.S.S. collaboration and reported that the gamma-ray emission from RX J1713.7À3946 arises mainly in the shell.

The observation of the source in the GeV energy region by Fermi telescope suggests a hard photon spectrum with power law spectral index $1.50 \pm 0.11$. The overall GeV to TeV energies can be explained by cosmic ray interactions with ambient matter assuming a very hard spectrum of protons with power-law index $1.7$ and an exponential cutoff at 25 TeV. The estimated upper cut-off (at 25 TeV) raises doubt on the acceleration of cosmic ray particles to PeV energies by RX J1713.7–3946 SNR.   
    
The distance of the SNR from the Earth is $\sim 1$ kpc and the radius of the shell is about 10 pc. The ambient matter density of the SNR is $\sim 1$ cm$^{-3}$ [38]. The total mechanical explosion energy of the supernova is taken as $E_{sn} = 1\times 10^{51}$ eV.

\begin{figure}[h]
  \begin{center}
\includegraphics[scale=0.44]{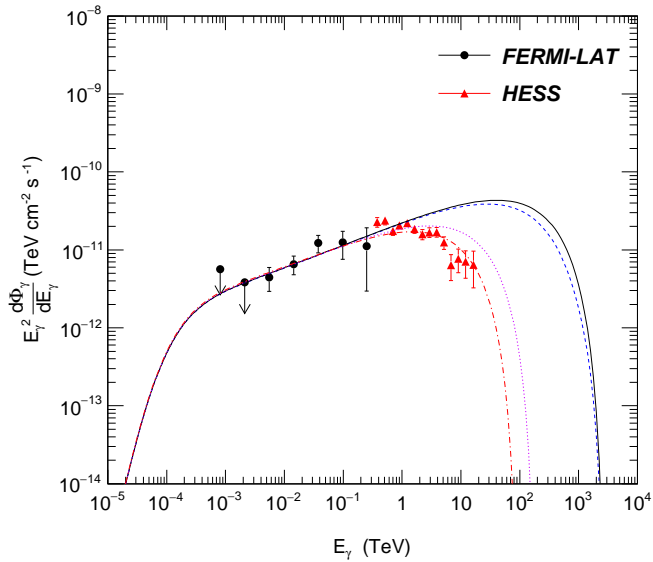}
\end{center}
\caption{Same as Fig.1 but for SNR RX J17137-3946. The red dash-doted line denotes the gamma ray flux considering maximum attainable energy of cosmic ray is 100 TeV}
\label{Fig:4}
\end{figure}

The calculated high energy gamma ray flux from the SNR reaching at earth is given in Fig. 4. The observed data can be explained well assuming the spectral index of the energy spectrum of SNR accelerated cosmic rays $\alpha = -1.8$. When we consider proton as a primary cosmic ray spectrum up to $100$ TeV, $15\%$ of the total explosion energy is needed in accelerated particles to fit the FERMI-LAT and HESS observational data. If primary cosmic ray proton up to $200$ TeV is considered then $17\%$ efficiency of such energy conversion is required. Again if primary cosmic ray proton up to 3 PeV is considered then $30\%$ efficiency of such energy conversion is needed. For the mixed primaries with rigidity dependent cut-off, $32\%$ and $22\%$ efficiency of conversion are needed for nuclear mass enhancement factor $A^{3/4}$ and $A$ respectively.

\subsection{IC 443}
Gamma ray emission from two middle aged SNRs, IC443 and W44 are detected over sub-GeV to TeV energies. The observation of spectral continuum down to 200 MeV from these two sources often attributed to a neutral pion emission. Both are asymmetric shell type SNRs. The gamma ray emission from W44 comes from two regions of the SNR, which are likely to be embedded molecular cluster, but not from the entire SNR. We, therefore, choose IC 433 here. 

IC 443 is located off the outer Galactic plane nearly at a distance of about 1.5 kpc [42]. It is a strong x-ray source. The EGRET telescope first detected gamma ray flux from the source above 100 MeV [43]. The Major Atmospheric Gamma Imaging Cerenkov (MAGIC) telescope first detected TeV gamma rays from IC 433 [44]. Later Fermi observed the source in the GeV energy range [45] and the VERITAS confirmed the TeV gamma emission from the SNR [46]. The MAGIC detection is displaced towards south from the EGRET source and it was argued that MAGIC detected TeV emission from IC 433 essentially comes from a giant cloud in front of the SNR [47].

\begin{figure}[h]
  \begin{center}
\includegraphics[scale=0.44]{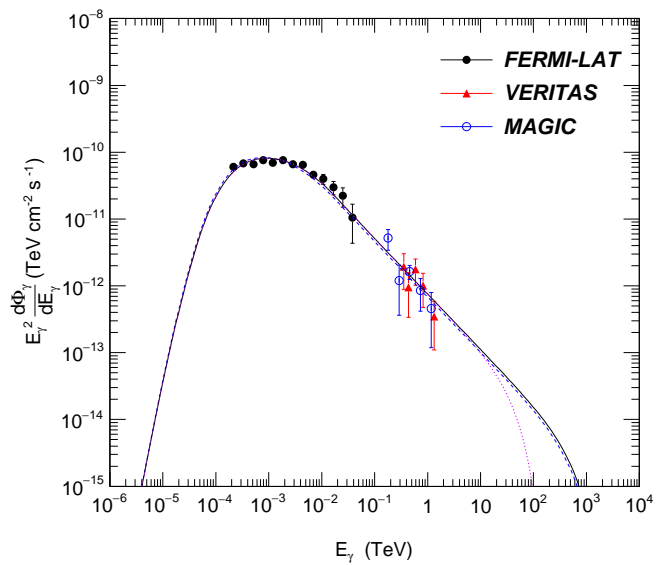}
\end{center}
\caption{Same as Fig.1 but for SNR IC 443}
\label{Fig:5}
\end{figure}

The explosion energy is not clearly known by any other means. Hence we take the standard value $E_{sn} = 1\times 10^{51}$ eV. The molecular environment suggests $n_{H} = 20 cm^{-3}$ [8].  The estimated high energy gamma ray flux from the SNR reaching at earth following Eqs. (2) - (10)  is displayed in fig. 5 along with the Fermi and MAGIC observed data points. Here we consider broken power law for the SNR accelerated cosmic ray energy spectrum and it is found that spectral index $\alpha = -2.1$ below 30 Gev and $\alpha = -2.9$ above it fits the observed data well. A 10\% efficiency of conversion is needed  for pure accelerated protons to explain the observed gamma ray spectrum. Whereas 17.5\% and 15.5\% efficiencies of conversion are required for the mixed composition with nuclear enhancement factor $A^{3/4}$ and $A$ respectively to explain the observed spectrum. 

\section{Discussion}
    
The main effect of considering cosmic ray nuclei with right abundances instead of pure proton on the secondary gamma ray spectra is the need of higher conversion efficiency. The conversion efficiencies required to match the gamma ray spectra of each of the SNRs considered here is shown in Table 1. Note that the SNR energy output in the Galaxy can supply the energy budget required to maintain the present population of cosmic-rays if the overall efficiency of conversion of explosion energy into cosmic ray particles is of the order of $10\%$. Nearly the same conversion efficiency is required to explain the high energy gamma ray emission observed from different SNRs in the galaxy in terms of interactions of SNR accelerated protons with the ambient matter and therefore the scenarios (cosmic ray density and gamma ray emission from SNRs) are mutually consistent. However, when mixed composition is invoked in evaluating the gamma ray spectrum, a higher efficiency is needed. Such higher conversion efficiency seems also necessary to maintain the observed cosmic ray energy density over a long period as the gamma ray observations already indicate that all the SNRs are not the generators of hadronic cosmic rays. 

\begin{table*}[t] 
  \begin{center} 
    \caption{Model fitting parameters for the SNRs where $\xi_{1}$ is measured considering wounded nuclei approach and $\xi_{2}$ is measured considering nuclear enhancement approach. } 
    \label{tab:table1} 
    \begin{tabular}{|m{2.3cm}|c|c|c|c|c|c|c|c} 
      \toprule 
      Supernovae & Model & Composition & $E_{cut}$ (eV)& $E_{b}$ (GeV)&$\alpha$ & $\xi_{1}$ ($\%$) & $\xi_{2}$ ($\%$)    \\ \hline 
      Cas A& BPL & p & $2\times 10^{14}$ &  &$\alpha_{1} = -1.7$, & $10$& $10$ \\  
           &     & p & $3\times 10^{15}$ & $50$ & $\alpha_{2} = -2.45$ & $10$&  $10$\\ 
           &     & Mixed & $Z\times3\times 10^{15}$ &  & & $16.5$& $14$ \\ \hline 
      Tycho & SPL & p & $2\times 10^{14}$ &  & & $12$&  $12$ \\ 
           &     & p & $3\times 10^{15}$ & $-$  &$-2.3$ & $12$&  $12$\\ 
           &     & Mixed & $Z\times3\times 10^{15}$ &  & & $19.8$&  $16.8$ \\ \hline 
      SN 1006 & SPL & p & $2\times 10^{14}$ &  & & $10$&  $10$\\  
           &     & p & $3\times 10^{15}$ &$-$ & $-2.05$  & $11.5$& $11.5$ \\ 
           &     & Mixed & $Z\times3\times 10^{15}$ &  & & $16$& $12.5$ \\ \hline 
      RX J1713.7−3946 & SPL & p & $1\times 10^{14}$ &  & & $15$& $15$ \\  
           &     & p & $2\times 10^{14}$ &    &       & $17$ & $17$  \\ 
           &     & p & $3\times 10^{15}$ & $-$ &$-1.8$ & $30$& $30$   \\ 
           &     & Mixed & $Z\times3\times 10^{15}$ &  & & $32$& $22$ \\ \hline 
      IC 443 & BPL & p & $2\times 10^{14}$ & & $\alpha_{1} = -2.1$,  & $10$ & $10$ \\  
           &     & p & $3\times 10^{15}$ &  $30$ &$\alpha_{2} = -2.9$   & $10$ & $10$ \\ 
           &     & Mixed & $Z\times3\times 10^{15}$ &  & & $17.5$& $15.5$ \\ \hline 
    \end{tabular} 
  \end{center} 
\end{table*}

A point to note that the slope of spectra of accelerated cosmic rays required to explain the observed gamma ray spectra in different SNRs is not unique. It is important to understand how such energy spectra of SNR accelerated cosmic rays with deviating spectral slopes lead to  cosmic ray energy spectrum with a universal spectral slope. 
    
In recent theoretical developments of the diffusive shock acceleration theory in SNRs, it is argued that a significant amplification of the magnetic field occurs as a result of the pressure gradient of the accelerating cosmic rays and thereby protons might be accelerated in SNRs up to the knee energy of the spectrum whereas heavier nuclei will have Z times that of protons. In such a scenario the gamma ray flux from young SNRs would be significantly higher at few tens of TeV and higher energies than the flux corresponds to cosmic rays with maximum energy limited to 200 TeV or so. The next generation gamma ray telescopes like CTA should be able to discriminate these two scenario of maximum energy.    
There is a possibility that the higher energy cosmic ray particles already might have escaped from the SNRs considered here but it is quite unlikely that even for the young SNR Cas A such leakage happens already, since in the standard DSA scenario, particles up to PeV energies are likely to be confined in the remnant over a period of $10^{4}$ years or so [3,48]. 

\section{Conclusion} 

There is now broad consensus that bulk of the cosmic rays with energies at least up to the second knee are originated in galactic SNRs where they are accelerated by diffusive shock acceleration process at supernova blast waves driven by expanding SNRs. It is now also generally believed that higher energy ($> 200$ TeV) particles are accelerated in the early phases of the supernova explosion (i.e. in young SNRs) though so far there is no experimental support in favor of this SNR paradigm. Further, if SNRs are true generators of galactic cosmic rays, they should accelerate not only protons but different cosmic ray nuclei with proper abundances. The gamma rays those produced in interaction of SNR accelerated cosmic rays with ambient matter may contain the imprints of such features (acceleration of cosmic ray nuclei to PeV energies). To explore such signatures in this work we estimate the hadronically produced high energy (GeV-PeV) gamma rays to be emitted by individual SNRs considering that i) SNRs accelerate cosmic ray particles with mass composition consistent with the observed mass composition of cosmic rays and ii) the maximum attainable energy of cosmic rays in SNRs is $Z \times 3 \times 10^{15}$ eV which is needed to explain  the cosmic ray spectrum till $100$ PeV including the knee and the second knee features. Comparing with the observations we evaluate the conversion efficiency and we also obtained the gamma ray spectrum till PeV energies which is beyond the upper energy limit of detection of the presently operating telescopes but within the reach of the forthcoming gamma ray telescopes like CTA. 
     
The nature of gamma ray emission spectra from SNRs is found almost independent of type of SNR accelerated cosmic ray nuclei if the spectral slope of each nuclei species is taken as the same. The energy spectra of cosmic ray heavier nuclei are harder than that of cosmic ray protons. If such a feature is adopted, the SNR produced gamma ray spectrum is expected to be a slightly harder than what we got. An interesting point is that to match the high energy gamma ray spectrum from individual SNRs with SNR accelerated cosmic ray nuclei instead of pure proton, the conversion efficiency has to be taken nearly double ($\sim 20 \%$) in comparison to those produced by pure proton cosmic rays. A conversion efficiency of the order of $20 \%$ is not unrealistic, but of course is more demanding. The density of SNR ambient matter and the total explosion energy are two important parameters in our estimation of gamma ray flux. We have taken the values those obtained by previous authors from different consideration. But still some uncertainties remain on these parameters and thereby absolute value of efficiencies are also somewhat uncertain.   

Regarding the issue of maximum energy of cosmic rays in SNRs, we compare between two different scenarios; $2 \times 10^{14}$ eV which is the theoretical upper limit under normal magnetic field picture and $Z \times 3 \times 10^{15}$ eV, which seems achievable under amplified magnetic field situation. The later (Pevatron) scenario is, in fact, essential for SNR origin model of galactic cosmic rays. We find that both the scenarios somewhat can describe the observed gamma ray spectra of all the SNRs considered in this work except RX J17137-3946. In the case of RX J17137-3946, the maximum cosmic ray energy appears to much lower and it is more likely that gamma ray emission from RX J17137-3946 is leptonic in origin. The Pevatron scenario in fact better describes the TeV gamma ray observations from SNR tycho and SN1006. The stated two scenarios give significant different fluxes above few tens of TeV and therefore the experiment HAWC or the upcoming experiments like CTA should be able to discriminate the two maximum energy pictures. 

\section*{Acknowledgments}
The authors would like to thank an anonymous reviewer for encouraging comments and pointing out a couple of typographical/latex errors.

\end{document}